%% file: acts.tex
\documentclass[10pt, paper=a4, UKenglish]{article}
\usepackage{graphicx}
\def\Title#1{\begin{center} {\Large #1 } \end{center}}
\def\Author#1{\begin{center}{ \sc #1} \end{center}}
\def\Address#1{\begin{center}{ \it #1} \end{center}}

\newcommand\pubblock{\rightline{\begin{tabular}{l} Proceedings of CTD 2020\\ \pubnumber\\
         \pubdate  \end{tabular}}}

\newenvironment{Abstract}{\begin{quotation} \begin{center} 
             \large ABSTRACT \end{center}\bigskip 
       \begin{large}}{\end{large} \end{quotation}}

\newenvironment{Presented}{\begin{quotation} \begin{center} 
             PRESENTED AT\end{center}\bigskip 
      \begin{center}\begin{large}}{\end{large}\end{center} \end{quotation}}

\def\Acknowledgements{\bigskip  \bigskip \begin{center} \begin{large}
      \bf ACKNOWLEDGEMENTS \end{large}\end{center}}

\input econfmacros.tex

\usepackage{color}
\usepackage{lineno}
\usepackage{subfig}
\usepackage{hyperref}

\newcommand\pubnumber{PROC-CTD2020-54}

\newcommand\pubdate{\today}

\def\affiliation{
On behalf of the ACTS developers, \\
Department of Physics \\
University of California, Berkeley CA, United States of America}

\def\support{\footnote{Work supported by the National Science Foundation}}

\newcommand{\conference}{Connecting the Dots Workshop (CTD 2020)\\
April 20-30, 2020}

\usepackage{fancyhdr}
\definecolor{mygrey}{RGB}{105,105,105}
\begin{document}


\large
\begin{titlepage}
\pubblock

\vfill
\Title{Tracking with A Common Tracking Software}
\vfill

\Author{Xiaocong Ai \support}
\Address{\affiliation}
\vfill

\begin{Abstract}
In high energy physics (HEP) experiments, the reconstruction of charged particle trajectories is one of the most fundamental yet computationally expensive parts of event processing. At future hadron colliders such as the High-Luminosity Large Hadron Collider (HL-LHC), there can be up to ten thousand particles per event. This increases the track reconstruction time by a factor about 5 compared to the current tracking environment. Efficient and fast tracking software is necessary to maintain and improve the tracking performance. This can benefit from both fast tracking algorithms and modern computing architectures with many cores and accelerators.

The \texttt{Acts} (A Common Tracking Software) project encapsulates the current ATLAS tracking software into an experiment-independent software designed for modern computing architectures. It provides a set of high-level track reconstruction tools agnostic to the details of the detector and magnetic field configuration. Particular emphasis is placed on thread-safety of the code in order to support concurrent event processing with context-dependent detector conditions, such as detector alignments or calibrations.
\texttt{Acts} aims in addition to be a research and development platform for studying innovative tracking techniques and exploiting modern hardware architectures.
\end{Abstract}

\vfill

\begin{Presented}
\conference
\end{Presented}
\vfill
\end{titlepage}
\def\thefootnote{\fnsymbol{footnote}}
\setcounter{footnote}{0}
%

\normalsize 


\section{Introduction}
\label{intro}
The Large Hadron Collider (LHC), operated at CERN, Geneva (Switzerland) will go through an ambitious series of upgrades that will ultimately result in an instantaneous luminosity up to 
$L = 7.5 \times 10^{34} \texttt{cm}^{-2} \texttt{s}^{-1}$ at the HL-LHC~\cite{HL-LHC}. This corresponds to approximately 200 inelastic proton-proton collisions per beam crossing (pileup, $\mu$) and will present an extremely challenging tracking environment to the existing ATLAS~\cite{ATLAS} tracking software~\cite{atlas-track-sw}, which was initially designed without much consideration of thread-safety. In order to facilitate this evolution, a redesigning as a new project rather than in situ changes of the ATLAS tracking software has been chosen. 

The \texttt{Acts} (A Common Tracking Software) project~\cite{acts1, acts2} aims to prepare a tracking toolkit for future detectors based on the ATLAS tracking software, and in addition open the software stack to inclusion of alternative ideas or concepts. In order to take advantage of modern CPU architectures with many parallel threads, the software is designed to be thread-safe by adhering to strict discipline of \texttt{const-correctness} and utilizing stateless tools. The code is written in compliance with the C++ 17 standard. 
To support fully parallel event execution, the context-dependent data,~i.e. detector alignment, calibration data, detector or magnetic field status, are handled by context or payload objects which guarantee access to the correct conditions in memory in a concurrent environment.
More importantly, \texttt{Acts} handles these context objects through the call chain,~i.e. \texttt{Acts} doesn't really implement them (the experiments do), but \texttt{Acts} makes sure the context objects arrive where they should.
In \texttt{Acts}, virtual interfaces are minimized by using compiler templating, which provides fast code execution and maximum customizability. \texttt{Acts} is designed with minimal dependency on external libraries with the \texttt{Eigen} library~\cite{eigen} as the only requirement. This vastly facilitates the integration of \texttt{Acts} into various experiment software frameworks.
Benefitting from its highly modular and user-friendly design, \texttt{Acts} is also used as a research and development (R\&D) platform for investigating various innovative tracking techniques and modern hardware architectures. 

The current release of \texttt{Acts} is v0.26.00. The main component of the repository~\cite{acts-core} is a set of fundamental tracking tools,~e.g. geometry and navigation, tracking Event Data Model (EDM), track parameter propagation engine and tracking algorithms. It also includes a \texttt{Gaudi}~\cite{gaudi}-inspired test framework which can be used to run event processing with prototype detectors to test the performance of the core components. In addition, a fast simulation engine used to simulate the trajectories of particles in the detector is provided in the repository.

This paper will present a brief overview of the \texttt{Acts} project. Section~\ref{components} will discuss selected components of the repository focusing on the tracking infrastructures and algorithms for track fitting and track finding. Section~\ref{performance} will demonstrate the performance of implemented algorithms for track fitting and track finding with an example detector. The ongoing R\&D activities in \texttt{Acts} will be introduced in Section~\ref{RD}. The conclusions will be provided in Section~\ref{conclusions}. 

\section{\texttt{Acts} tracking components}
\label{components}

\subsection{Geometry and navigation}
The detector geometry is the fundamental component for track parameter propagation and integration of material effects during track reconstruction. Since track reconstruction with a realistic detector description as used in full detector simulation requires large CPU, a simplified detector geometry,~i.e. tracking geometry, is used in track reconstruction in \texttt{Acts} for efficient navigation and fast extrapolation of tracks.

The key geometric component of the \texttt{Acts} tracking geometry is the \texttt{Acts::Surface} class, which builds up into further geometrical objects such as the \texttt{Acts::Layer} and \texttt{Acts::TrackingVolume}. 
The \texttt{Acts::Surface} class plays the key role for defining the local reference frame for track representation. It is fundamental to the tracking EDM and used together with track parameter propagation. In order to support various detector layouts, the abstract \texttt{Acts::Surface} base class is inherited by various concrete surface types with different definitions of local reference frame: the \texttt{Acts::PlaneSurface} with a local Cartesian coordinate system, the \texttt{Acts::DiscSurface} with a local polar coordinate
system, the line-like surfaces for describing straw detectors (\texttt{Acts::StrawSurface}) and perigee representation (\texttt{Acts::PerigeeSurface}) as well as the \texttt{Acts::CylinderSurface} and the \texttt{Acts::ConeSurface} with appropriate cylindrical coordinates. 
The shapes of the surfaces are described by the inherited concrete classes of the \texttt{Acts::SurfaceBounds}.
The \texttt{Acts::TrackingVolume} contains \texttt{Acts::BoundarySurface} objects inherited from the \texttt{Acts::Surface} class with additional information about the attached \texttt{Acts::TrackingVolume}. Its oriented normal vector is used as an indicator of the track propagation direction, which turns the \texttt{Acts::BoundarySurface} into the key component for navigation across the \texttt{Acts::TrackingVolume}. 

Each concrete surface type can be approximated by an \texttt{Acts::Polyhedron} object which contains a list of Cartesian vertices in the global frame as well as a list of indices indicating the vertices to form a face. This polyhedron approximation of the surface allows for a common representation of the various geometrical objects contained in the \texttt{Acts::TrackingGeometry}. As a consequence, the intersection of a track with all geometrical objects can be investigated simultaneously harnessing the massive parallelism of modern hardwares. 

The detector material approximated from the full detector geometry description is mapped into the tracking geometry,~i.e. the \texttt{Acts::Surface} or the \texttt{Acts::Volume}, by a dedicated material mapping algorithm. When the material is mapped into the \texttt{Acts::Surface}, the material effects are considered when the surfaces are traversed by the track. When the material is mapped into the binned grids of the \texttt{Acts::Volume}, the material effects from each grid are considered when the track propagation reaches that grid.

The full detector geometry description is based on the abstract \texttt{Acts::DetectorElementBase} class with a pure virtual method to return the \texttt{Acts::Surface} representation, which helps build the connection between the full detector geometry and the \texttt{Acts} tracking geometry. \texttt{Acts} has dedicated geometry plugins to build full detector geometries from various detector geometry description,~e.g. the TGeo within the \texttt{ROOT Geometry Package}~\cite{tgeo} and the common DD4Hep modelling~\cite{dd4hep}. To date, the following HEP detector types have been implemented in the \texttt{Acts} tracking geometry: the Silicon tracker, the Time Projection Chamber (TPC), the Calorimeter and the Muon Chambers.  
 
\subsection{Event Data Model}
In \texttt{Acts}, the EDM used for track reconstruction is based on the \texttt{Eigen} library.
A track can contain one or more sets of track parameters along the track and the measurements from which the track parameters are derived. The available tracking information on a surface is grouped together into a track state object. Thus, the track state EDM and track EDM are usually defined based on the track parameter EDM and measurement EDM.  

The trajectory of a particle in a magnetic field can be described by a set of track parameters describing the space coordinates and momentum of the particle at that space point. In \texttt{Acts}, an additional time parameter is also included to support inherent timing measurement. The global or free track parameters are represented by the following set of variables~\footnote{A helical representation has deliberately not been chosen, as this would bind the parameterization to a solenoid field}:
\begin{equation}
 G = (\vec{X}, t, \hat{T}, \frac{q}{p}),
\end{equation}
where $(\vec{X}, t)$ represents the coordinate in space-time, $\hat{T}$ an unit vector representing the direction of global momentum, $q$ the charge and $p$ the momenta of the track. The local representation of the track parameters is:
\begin{equation}
 L = (l_0, l_1, \phi, \theta, \frac{q}{p}, t), 
\end{equation}
where the $(l_0, l_1)$ denotes two coordinates to describe the position in the local frame, $(\phi, \theta, \frac{q}{p})$ a representation of the momentum in the three-dimensional polar frame and $t$ the timing component of the space-time coordinate. Two types of local track parameter EDM are implemented and used in different scenarios: the \texttt{Acts::SingleBoundTrackParameters} class when the reference surface is an explicit geometry object and the \texttt{Acts::SingleCurvilinearTrackParameters} class when the reference surface is implicitly built as an \texttt{Acts::PlaneSurface} with its center defined at the current track position and its normal vector pointing along the momentum direction. The $(l_0, l_1)$ of an \texttt{Acts::SingleCurvilinearTrackParameters} object is fixed to be $(0, 0)$. 
For the \texttt{Acts::SingleBoundTrackParameters} class, the $(l_0, l_1)$ can represent local coordinates with different parameterization depending on the type of its reference surface. For example, the $(l_0, l_1)$ represents the local position in the Cartesian coordinate when the reference surface is an \texttt{Acts::PlaneSurface} object and represents the two dimensional polar coordinates when the reference surface is an \texttt{Acts::DiscSurface}.
An appropriate local coordinate system is chosen to allow an ideal mapping between measurement representation and track parameterization and to minimize coordinate transformations.

The \texttt{Acts::Measurement} class is a templated class that receives the original raw data as a template parameter. It contains a variable set of measurement parameters from the full track parametersiation determined at compile time. This is to achieve maximum performance. The design of the \texttt{Acts::Measurement} class allows for user-defined measurement definition which is usually detector-specific as well as an additional calibration correction to the original measurement during the track reconstruction.

The \texttt{Acts::TrackState} class is designed based on the concept of the Kalman Filtering~\cite{kalman} algorithm,~i.e. it contains three optional sets of track parameters representing the predicted, filtered and smoothed track parameters from the fitting, the optional uncalibrated measurement,~i.e. a link to the raw detector measurement, and the optional calibrated measurement with user-defined calibration correction, at the reference surface. The \texttt{Acts::TrackState} also contains additional information for the transport Jacobian from the previous track state, the Kalman filtering quality and a bit-set type flags for fast distinction between different types of track state,~e.g. whether the measurement contained in the track state is an outlier (measurement which is incompatible with the track hypothesis). 

The combinatorial Kalman Filtering usually requires a tree-like structure for the track states, therefore, a dedicated track EDM, the \texttt{Acts::MultiTrajectory} class, is designed as a container of a collection of track states. For an \texttt{Acts::MultiTrajectory} object, the different types of information contained in the \texttt{Acts::TrackState} are stored separately. Bookkeeping of the storage indices uses a dedicated index data type. In particular, the index of the previous track state is stored in the index data of the current track state. This allows for multiple branches with a common previous track state, which is inherently consistent with the concept of the combinatorial Kalman Filtering. A track state proxy helps read from and write to the storage for a track state with a specific index. 

\subsection{Track parameter propagation}
The extrapolation of the track parameterization and associated covariances is an integral part of track reconstruciton. In general, this involves the (mostly) numerical integration of the equation of motion. In \texttt{Acts}, this is done with the \texttt{Acts::Propagator} class, which can use different numerical integrators.
The adaptive Runge-Kutta-Nyström method~\cite{ARKN} is used as the primary integration method. 
An extension of the standard Runge-Kutta-Nyström numerical integrator\cite{RKN}: a concept that has been pioneered by ATLAS prior to the LHC Run-1~\cite{STEP}, has been adapted in \texttt{Acts} as an appropriate propagator extension for propagation in dense material environment (e.g. Calorimeters). This extension is needed for aforementioned volume based material integration. 
The \texttt{Acts::Propagator} class also supports navigation through different tracking geometries. The \texttt{propagate} call requires a set of starting track parameters, the propagation options with a compile-time extendable list of \texttt{Actor} and \texttt{Aborter} to support execution of custom code and usage of custom abort conditions at each integration step and an optional target surface which can be used to construct an \texttt{Aborter} to stop the propagation and extract the track parameters. 

\subsection{Tracking algorithms}
Using the measurements as input, track reconstruction includes both the process of track finding, which uses either local or global pattern recognition methods to identify measurements originating from the same charged particle, and track fitting, which aims to estimate the track parameters from a set of measurements. Using the fitted tracks, the task of vertex reconstruction includes the identification of sets of tracks originating from the same vertex and the estimation of the parameters of the vertex,~e.g. position and covariance of the vertex. In \texttt{Acts}, various primary vertex reconstruction tools are available,~e.g. the \texttt{Acts::IterativeVertexFinder} and the \texttt{Acts::MultiAdaptiveVertexFinder}~\cite{vertex}.

\subsubsection{Tracking fitting}
The Kalman Filtering technique is implemented as the \texttt{Acts::KalmanFitter} in \texttt{Acts}. It is designed to be agnostic to the representation of the track parameters and measurements, the magnetic field, the detector geometry, the Kalman filtering and smoothing technique, the outlier finder and the calibrator for calibration of uncalibrated measurement using predicted track parameters during the fitting. It contains a properly configured \texttt{Acts::Propagator} which is served with a dedicated Kalman \texttt{Actor}. The core component of the \texttt{fit} method includes a \texttt{propagate} call. The \texttt{Actor} includes a method to create a track state if the propagation reaches a surface with either material or measurement on the surface. The found measurement is investigated by the outlier finder. Unless the measurement is tagged as an outlier by the outlier finder, it is used to update the track parameters by running the Kalman filtering. A hole track state is created on a traversed sensitive surface without measurement on it. The material effects can be included either before or after the Kalman filtering, which allows for splitted material effects. 

When all the measurements have been processed or the navigation reaches the boundary of the tracking geometry, the Kalman smoothing is triggered to get the smoothed track parameters by either using the Rauch-Tung-Striebel smoother~\cite{RTS} formalism starting from the last filtered track state or proceeding with the propagation but with the navigation direction reversed.

The \texttt{fit} method of the \texttt{Acts::KalmanFitter} returns an \texttt{Acts::MultiTrajectory} object containing only one fitted track and the fitted track parameters at a target surface specified via the \texttt{Acts::KalmanFitterOptions}.

\subsubsection{Tracking finding}
The track finding algorithms in \texttt{Acts} includes a track seeding algorithm, the \texttt{Acts::Seedfinder}, and a track following algorithm, the \texttt{Acts::CombinatorialKalmanFilter}, using the combinatorial Kalman Filtering technique. 

After transforming the measurements into space points (these are created from either a single two dimensional measurement, or a combination of two one dimensional measurements), the space points are grouped into two dimensional grids binned by the azimuthal angle and the global z coordinate of the space point. For each space point, the \texttt{Acts::Seedfinder} searches for all compatible space points from the neighbor grids by comparing the azimuthal angle to form the space point doublets, and then compare two doublets to search for a triplet by comparing the radius of the three space points in the transverse plane and the polar angle between the doublet.
Since the search for triplets for different space points is independent on the search order, the \texttt{Acts::Seedfinder} is highly parallelizable. The \texttt{Acts::Seedfinder} is also designed to be highly configurable allowing for flexible seed finding criteria driven by~e.g. potential minimum transverse momentum of the tracks, measurement errors and region of interest for seed finding.

As an extension of the \texttt{Acts::KalmanFitter}, the \texttt{Acts::CombinatorialKalmanFilter} can perform the measurement search during the fitting. If there are multiple compatible measurements found on a surface, the track propagation is repeated with multiple sets of track parameters updated with each measurement. The search of compatible measurements is handled by a measurement selector which supports custom implementation of the selection criteria. The \texttt{findTracks} method of the \texttt{Acts::CombinatorialKalmanFilter} returns an \texttt{Acts::MultiTrajectory} object with one or multiple tracks stored inside and multiple sets of fitted track parameters at the user-defined target surface. 

\section{Tracking performance}
\label{performance}
The performance of the \texttt{Acts::KalmanFitter} and the \texttt{Acts::CombinatorialKalmanFilter} algorithms are investigated using both a single muon sample with the transverse momentum $p_{T}$ uniformly distributed between 100 MeV and 1 GeV and pseudorapidity $\eta$ uniformly distributed between -2.5 and 2.5, and a $t\bar{t}$ sample with pileup, $<\mu> = 200$, with a simplified detector implemented for the TrackML~\cite{trackML} challenge and at the ATLAS magnetic field.

\subsection{The \texttt{Acts::KalmanFitter} performance} 
The fitting resolution is studied by calculating the pull of each track parameter:
\begin{equation}
pull = \frac{v_{fit} - v_{truth}}{\sigma_{v}},
\end{equation}
where $v_{fit}$ and $\sigma_{v}$ are the value and error of the fitted track parameter, respectively. The $v_{truth}$ is the corresponding truth particle parameter.
If the track parameter and its error are estimated correctly, the distribution of pull values should have mean 0 and R.M.S. equal to 1. This assumes that the measurement mapping function can be approximated by the first-order linearization model or the Gaussian noise from the material effects is large enough to underweight the non-linearization effect of the measurement model. 

\begin{figure}[!htb]
  \centering
  \includegraphics[width=1\linewidth]{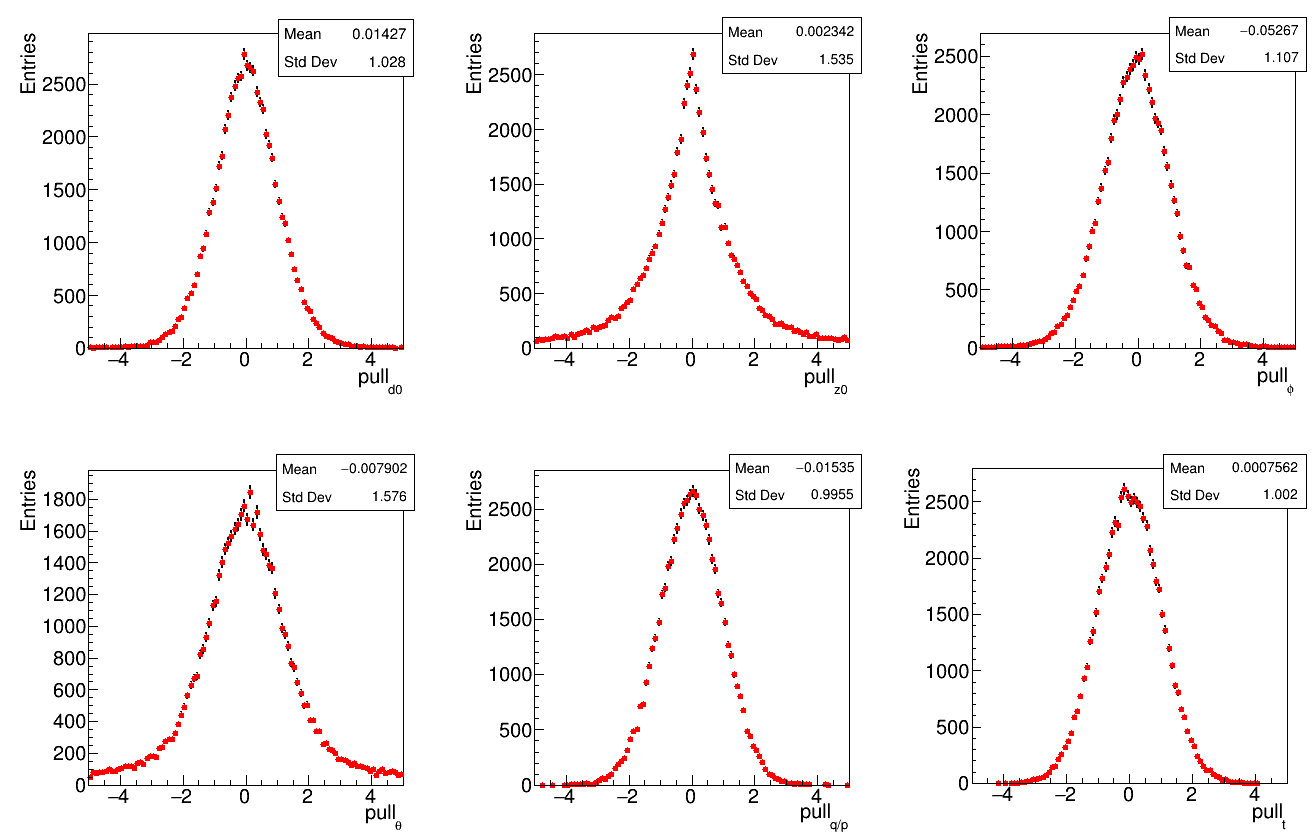}
  \caption{The distributions of pull values of the fitted perigee track parameters, $(d_0, z_0, \phi, \theta, \frac{q}{p}, t)$, for a sample of 1M single muons.
The R.M.S. of the pull distributions for the longitudinal impact parameter $z_{0}$ and the momentum direction $\phi$ and $\theta$ are deviated from value 1 since the non-linearization effect of the measurement model becomes significant with the TrackML detector which is modelled with little material.}
  \label{fig:pull_muon}
\end{figure}

The pull distributions of the set of fitted track parameters at a perigee surface defined with respect to the beam line (in this case, the component $(l_0, l_1)$ of the track parameters represents the impact parameters in the transverse and longitudinal plane,~i.e. $(d_0, z_0)$) using \texttt{Acts::KalmanFitter} for the single muon sample are shown in Figure~\ref{fig:pull_muon}. Since the TrackML detector is modelled with very little material, the non-linearization of the measurement model becomes significant, which results in some deviation of the R.M.S. of the pull distributions from the expected value 1. 

\begin{figure}[!htb]
  \centering
  \subfloat[]{\includegraphics[width=0.45\linewidth]{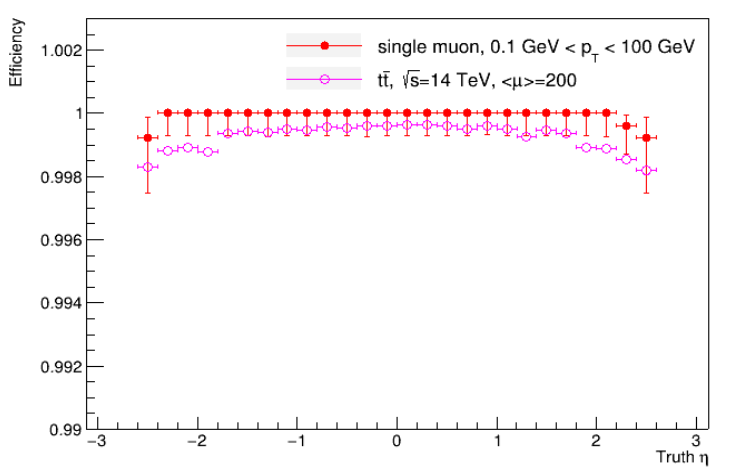}}
  \qquad
  \subfloat[]{\includegraphics[width=0.45\linewidth]{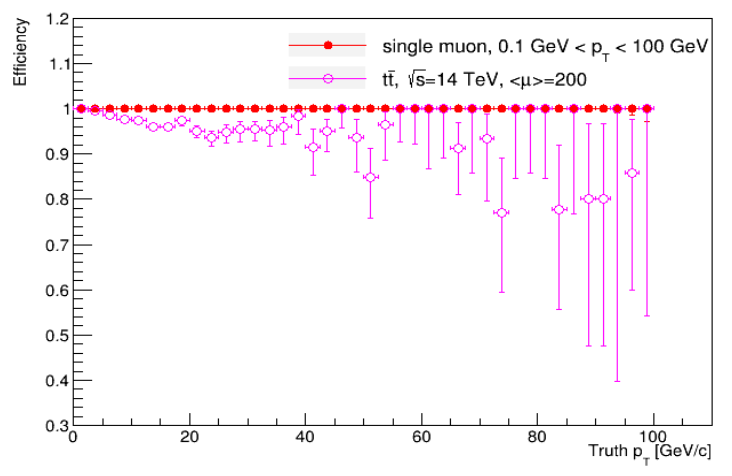}}
  \caption{The fitting efficiency versus (a) $\eta$ and (b) $p_{T}$ of the \texttt{Acts::KalmanFitter} for a sample of 1M single muons and 1k $t\bar{t}$ sample with $<\mu>$ = 200.}
  \label{fig:eff_kf}
\end{figure}

The fitting efficiency defined as the fraction of truth tracks that has fitted track parameters at the perigee surface versus $\eta$ and $p_{T}$ for both single muon and $t\bar{t}$ samples are shown in Figure~\ref{fig:eff_kf}. A technical fitting efficiency of 100\% is attained for the single muon sample. The slight efficiency loss for the $t\bar{t}$ sample is coming from those truth particles significantly displaced from the active detector region.  

\begin{figure}[!htb]
  \centering
  \includegraphics[width=0.7\linewidth]{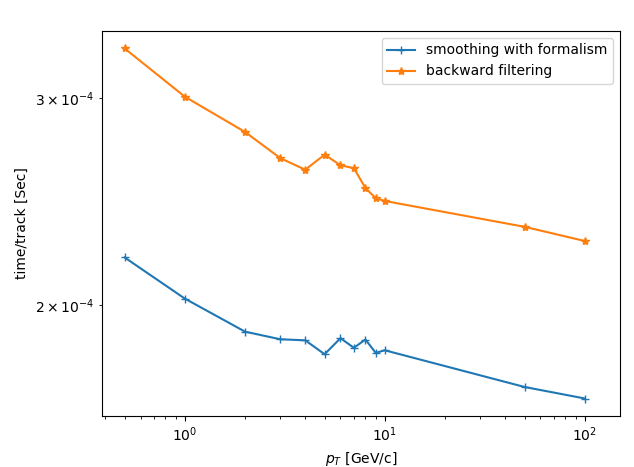}
  \caption{The average fitting time per track versus $p_{T}$ using the \texttt{Acts::KalmanFitter} tested with single muons at the Intel's i7-8559U processor.}
  \label{fig:kf_timing}
\end{figure}

Figure~\ref{fig:kf_timing} shows the average fitting time per track versus $p_{T}$ with the two different smoothing approaches tested with single muons at the Intel's i7-8559U processor with a maximum single core frequency at 4.5 GHz. With the Rauch-Tung-Striebel smoother used for the smoothing, an average fitting time, 0.2 ms per track, is achieved. This shows very good timing performance of the \texttt{Acts::KalmanFitter}.   

\subsection{The \texttt{Acts::CombinatorialKalmanFilter} performance}
Among the most important performance criteria for a track finding algorithm is the tracking efficiency and the rate at which fake tracks are reconstructed. 
The efficiency is usually defined as the fraction of particles which are associated with tracks passing a set of track quality requirements. The matching probability of a reconstructed track is defined as the fraction of measurements originating from the particle with the majority number of hits. The efficiency of the \texttt{Acts::CombinatorialKalmanFilter} is defined as the number of selected reconstructed tracks matched to a selected truth particle with a matching probability above 0.5, divided by the number of selected truth particles. The truth particles are required to have at least 9 hits on the detector and the reconstructed tracks are required to have at least 9 measurements.
The fake rate is defined as the fraction of reconstructed tracks that don't match to any truth particle,~i.e. have a match probability below 50\%. 

\begin{figure}[!htb]
  \centering
  \subfloat[]{\includegraphics[width=0.45\linewidth]{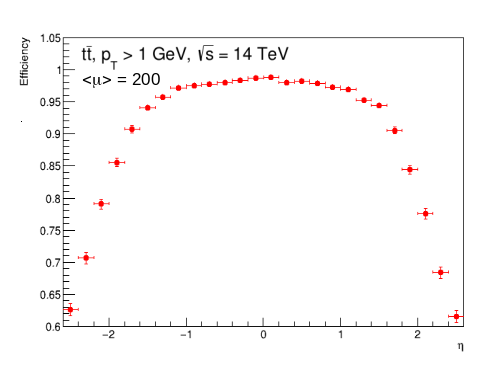}}
  \qquad
  \subfloat[]{\includegraphics[width=0.45\linewidth]{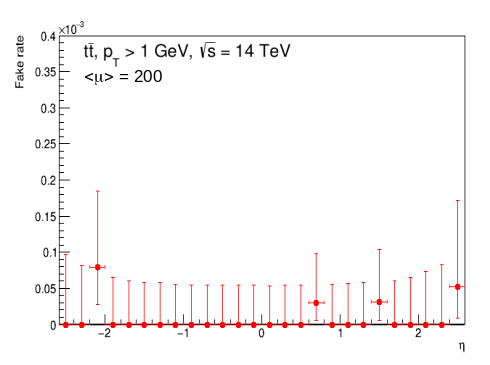}}
  \caption{The (a) efficiency and (b) fake rate of the \texttt{Acts::CombinatorialKalmanFilter} for 1k $t\bar{t}$ sample with $<\mu>$ = 200. Only truth particles with $p_{T}>1$ GeV are considered here.}
  \label{fig:perf_ckf}
\end{figure}

Figure~\ref{fig:perf_ckf} shows the efficiency and fake rate versus $\eta$ for the $t\bar{t}$ sample. The efficiency is about 99\% with central pseudorapidity and a low fake rate of $10^{-4}$ is attained. 

\section{R\&D with \texttt{Acts}}
\label{RD}
There are currently a number of projects exploring Machine Learning (ML) techniques with the goal of developing fast tracking algorithms,~e.g. the Kaggle TrackML challenge which has a range of solutions, the HEP.TrkX project using Graph Neural Networks~\cite{trkx} and the similarity hashing technique~\cite{hashing}. The TrackML detector implemented in \texttt{Acts} and the simulation data sets provided by the \texttt{Acts} fast simulation are widely used within those projects and many more. 

\begin{figure}[!htb]
  \centering
  \includegraphics[width=0.7\linewidth]{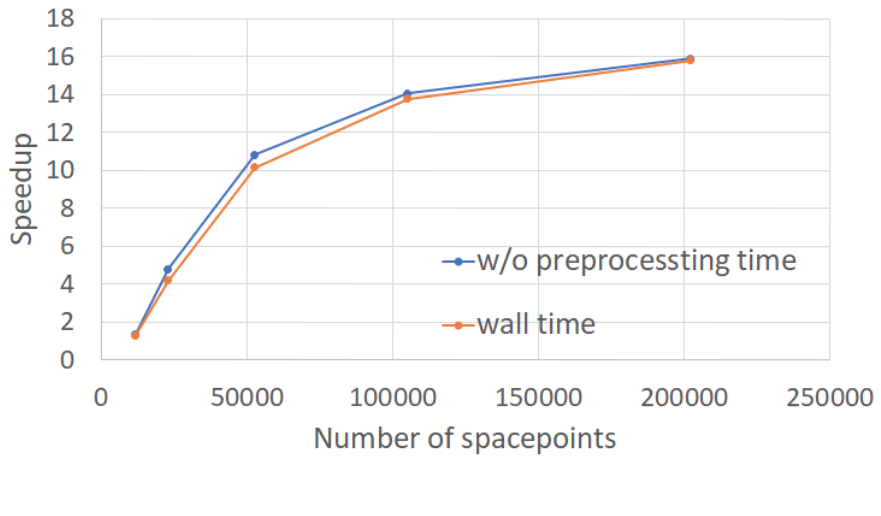}
  \caption{The gained speed-up which is defined as the ratio of seeding time on CPU (Intel's i7-5820K) and GPUs (NVIDIA GTX1070) versus number of space points for the \texttt{Acts::Seedfinder}. The orange line and blue line represent the speed-up with and without preprocessing time for the space points, respectively.}
  \label{fig:seedfinder}
\end{figure}

\begin{figure}[!htb]
  \centering
  \includegraphics[width=0.7\linewidth]{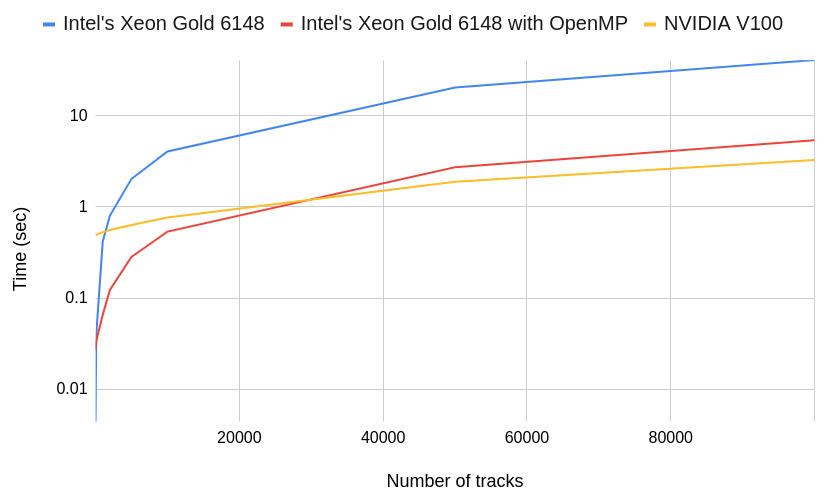}
  \caption{The propagation time with 1000 propagation steps using the \texttt{Acts::Propagator} at the ATLAS magnetic field versus number of tracks on CPU (Intel's Xeon Gold 6148) with (blue line) and without (red line) OpenMP parallelization, and on GPUs (NVIDIA V100) (yellow line).}
  \label{fig:prop}
\end{figure}

HEP will need to exploit modern hardware technologies with large number of cores and GPUs to speed up track reconstruction. Event level parallelization across many cores of the processor is key to gain speed. Using the \texttt{Acts::KalmanFitter}, intra-event level parallelization,~e.g. parallel fitting at the track level, with many cores has been investigated. Benefitting from the highly parallelizable structure of the \texttt{Acts::Seedfinder}, a factor of about 14 speed-up with identifcal physics output using GPUs compared to CPU at a tracking environment with up to 100k hits can be achieved as shown in Figure~\ref{fig:seedfinder} which shows the seeding time versus number of space points. The parallelization of track parameter propagation at the track level using GPUs has also been investigated. Figure~\ref{fig:prop} shows the propagation time using the \texttt{Acts::Propagator} at the ATLAS magnetic field versus number of tracks. The speed increase compared to CPU execution without any parallelization is observed.

\section{Conclusions}
\label{conclusions}
The anticipated large increase in track multiplicity at future colliders needs high performance tracking software and ability to exploit parallel architectures.
The \texttt{Acts} project aims to provide a framework-independent and detector-independent tracking toolkit tested for strict thread-safety
to support multi-threaded event processing.

The \texttt{Acts} package is designed to be highly applicable and customizable to meet the requirement of different experiments. It includes a state-of-the-art implementation of the tracking infrastructures,~e.g. tracking geometry, EDM and track parameter propagation engine and various tracking tools. In the past year, the tracking infrastructures are much optimized. In addition, new tools for track finding, tracking fitting and vertex reconstruction have been implemented.
The \texttt{Acts} project is currently undergoing very active development. In particular, there is rapidly growing interest in applying \texttt{Acts} as a tracking toolkit for various experiments over the past year,~e.g. the CEPC experiment~\cite{cepc}, the Belle-II experiment~\cite{belle-2}, the sPHENIX experiment~\cite{sphenix} and the EIC experiment~\cite{eic} and a number of detector geometries have been implemented. 

Ongoing projects within the community include simplifying the navigator structure, implementing the concept of tracking fitting with free track parameters and measurement representation and implementing of Kalman Filtering based alignment approach. In the future, we anticipate focus on optimizing and improving the tracking toolkit by vast application of \texttt{Acts} to detectors from a variety of experiments. 


\Acknowledgements
This work was supported by the National Science Foundation under Cooperative Agreement OAC-1836650.



\end{document}

%% file: econfmacros.tex
\def\beq{\begin{equation}}
\def\eeq#1{\label{#1}\end{equation}}
\def\eeqn{\end{equation}}

\def\beqa{\begin{eqnarray}}
\def\eeqa#1{\label{#1}\end{eqnarray}}
\def\eeqan{\end{eqnarray}}

\let\bar=\overbar

\def\Dslash{\not{\hbox{\kern-4pt $D$}}}
\def\dslash{\not{\hbox{\kern-2pt $\del$}}}

\def\msb{{\bar{\ssstyle M \kern -1pt S}}}

